# Who Gets to Come In? How Political Engagement Shapes Views on Legal Immigration

**Author:**
Muhammad Hassan Bin Afzal[1], Ph.D.
Visiting Assistant Professor
Department of Political Science and Public Service,
The University of Tennessee at Chattanooga
Hassan-Afzal@utc.edu

Foluke Omosun, Ph.D.
Assistant Professor
Department of Communication and Media
Sacred Heart University
Fairfield, Connecticut
omosunf@sacredheart.edu

**Abstract**
This study examines how political engagement shapes public attitudes toward legal immigration in the United States. Using nationally weighted data from the 2024 ANES Pilot Study, we construct a novel Political Engagement Index (PAX) based on five civic actions—discussing politics, online sharing, attending rallies, wearing political symbols, and campaign volunteering. Applying weighted ordered logistic regression models, we find that higher engagement predicts greater support for easing legal immigration, even after adjusting for education, gender, age, partisanship, income, urban residence, and generalized social trust. To capture the substantive effect, we visualize predicted probabilities across levels of engagement. In full-sample models, the likelihood of supporting "a lot harder" immigration drops from 26% to 13% as engagement rises, while support for "a lot easier" increases from 10% to 21%. Subgroup analyses by partisanship show consistent directionality, with notable shifts among Republicans. Social trust and education are also consistently associated with more open attitudes, while older respondents tend to support less easy pathways to legal immigration policies. These findings suggest that a cumulative increase in political participation is associated with support for legal immigration in shaping public attitudes toward legal immigration pathways, with varying intensity across partisan identities and socio-demographic characteristics.

**Keywords**
Political Engagement, Immigration Policy, Voting Behavior, Civic Participation, Issue Salience, Elite Cues

**This draft research paper was prepared for presentation at the MPSA 2025 Annual Conference and remains a work in progress. Please contact the corresponding author with any collaboration inquiries or questions about the study. We appreciate your interest.**

---

[1] Corresponding author, please contact the authors for any questions, thank you!

## Introduction

Issue saliency influences policy prioritization among the general public, often independent of personal experience or direct impact. In the context of the United States, the issue saliency of immigration and associated policies remains highly visible not necessarily because of widespread individual relevance but because public discussions in news media, campaigns, and institutional messaging consistently emphasize it as a topic of concern (Budge, 1982; Budge & Farlie, 1983; Zaller, 1992). As of January 2025, the U.S. labor market reported 7.7 million job openings, with healthcare, accommodation, and food services showing the highest vacancy rates (U.S. Bureau of Labor Statistics, 2025). Despite steady hiring at 5.4 million and 5.3 million separations, many roles remain unfilled, indicating structural demand. Immigration is often discussed as a solution to such labor gaps, yet public perceptions about immigration policy are shaped less by economic data and more by saliency (Alesina & Tabellini, 2024; Allen et al., 2024; Jin, 2024).

Prior research ties political engagement to issue salience, but findings remain mixed on whether increased political engagement dependably increases concern for specific issues like immigration or not (Hannuksela et al., 2024; Jennings & Wlezien, 2016; Paul & Fitzgerald, 2021). We build on this debate by constructing a unique Political Engagement Index (PAX) using five participatory items from the 2024 ANES Pilot Study[2], supported by strong scale reliability ($\alpha = .731$) and robust factor structure (RMSEA = .061, CFI = .983). We aim to explore the impacts of cumulative political engagement among individuals and their issue saliency about immigration and attitudes towards legal immigration; unlike prior studies where the political engagement is captured through as a single item or dichotomy, our index captures both online and offline political engagement behaviors, allowing a more nuanced test of whether political engagement shapes perceptions and saliency of immigration as a national priority. Therefore, our research study primarily focuses on how increasing cumulative political engagement shapes U.S. voters' perception of the importance of immigration policy. We aim to contribute to the growing scholarship on how issue attention is shaped by individual attitudes and political context, engagement, and information flow.

## Saliency Theory

Saliency theory focuses on how political parties make specific issues more salient so that the public can gain an advantage in elections (Budge, 2015). Political parties direct the public's focus toward issues that benefit their own interests and downplay others rather than focusing on the opposition's policies (Dolezal et al., 2014). Scholars have investigated the issue of salience in campaigns and elections and the effect on voters (e.g., Adams et al., 2023). While salience or importance is crucial, audience attention to these issues also plays a key role in impacting their attitudes and behaviors (Moniz & Wlezien, 2020). When the issues that parties are linked to rank high on the public agenda during political campaigns, they are advantaged because it likely leads to votes (Budge, 2015).

The concept of issue ownership also plays a vital role in influencing voters. A party's previous position or prior policy proposals often shape the public's perception of its ownership of those issues, and how the public views the party's credibility and competence on those specific issues—which can ultimately influence voter behavior (Budge, 2015; Williams & Ishiyama, 2022). The media in particular plays a significant role in drawing attention to important issues, which leads to them rising to the top of the public agenda (Czymara & Dochow, 2018). This amplification of issues is further supported when targeted political messaging and advertising through digital channels empowers parties to emphasize issues of importance to certain segments of the public (Chu et al., 2023). In other words, when politicians and parties amplify immigration issues, it makes it more important to the public (Paul & Fitzgerald, 2021). However, Jin (2024) noted the discrepancy in research findings, where some scholars have found that media salience regarding immigration links to anti-immigrant attitudes, while others have found no correlation, suggesting the need for further investigation.

---

[2] The ANES 2024 Pilot Study is a nationally representative, cross-sectional survey conducted by the American National Election Studies to test new questions and instruments ahead of the full 2024 Time Series study, offering early insight into voter attitudes, behaviors, and political values in the lead-up to the U.S. presidential election (ANES, 2024).

## Elite Cues

The elite cues framework has its origins in Zaller's (1992) work on how public opinions are formed specifically regarding political issues. Zaller holds that the public's political views are shaped by elites in government and media. The public depends on political elites to shape their political views rather than engage with the complexities of those issues on their own (Gilens & Murakawa, 2002).

Scholars have previously found that elite cues influence voter attitudes and behaviors. For example, Morisi and Leeper (2024) found that when individuals were informed that most political elites supported the UK's membership in the EU, their election predictions shifted to align with the stance of these elites. Similarly, elite cues were found to be a major factor for distrust in scientific experts during COVID-19 (Hamilton & Safford, 2021). Even when political elites hold positions that contrast with expert opinion, citizens often disagree with experts "when political elites they favor challenge this opinion" (Darmofal, 2005, (p. 381). However, the influence of elite cues on attitudes and behaviors depends on the context (Dickson & Hobolt, 2024; Jones & Ford, 2017). Specifically, Dickson and Hobolt (2024) highlighted the potential risks posed by elites who deliberately encourage actions that undermine democratic norms and institutions.

Elite cues are more likely to be effective when individuals perceive them to be similar to their own beliefs, when the issue is complex, or when they have low levels of involvement or political knowledge (Gilens & Murakawa, 2002). However, according to Gilens and Murakawa (2002), "while elite cues *can* provide an efficient shortcut to political decision making, the extent to which they are used and their effectiveness as a substitute for substantive knowledge remain unclear" (p. 43).

## Immigration Saliency, Political Engagement, and Public Opinion

Over the years, immigration has become a salient issue in U.S. and European elections (Harteveld et al., 2017; Jones & Ford, 2017). Immigration-related cues can have a powerful effect in shaping public opinion and policy (Jones & Ford, 2017). For example, in their study of German attitudes toward immigration, Czymara and Dochow (2018) found that when immigration was prominently discussed in the media, public concern increased. However, they also found that concerns were less likely for individuals living in areas with more immigrants. Issue salience can also influence the public's perception that other members of the public view immigration as an important issue (Hopkins, 2011).

Researchers found a shift in the U.S. over the years to a more restrictive position on undocumented immigration by the Republican Party, particularly post-9/11 (Gonzalez O'Brien et al., 2019). The U.S. is witnessing high levels of polarization (Kim & Rojas, 2025). Likewise, immigration in the U.S. is a polarizing topic that is steeped in partisan politics. Public concerns regarding economic, cultural, and security threats fuel negative attitudes toward immigrants (Hellwig & Kweon, 2016). "Stoking fears about cultural outsiders is a perennial strategy for motivating social groups who feel threatened to support right-wing populist politicians and punitive social policies targeting minorities" (Baker & Edmonds, 2021, p. 300).

Jin (2024) argued that when political elites disagree on immigration, anti-immigrant sentiments and attitudes are perpetuated, however, when political elites have consensus on the issue, there is neutrality toward the problem by the public. The two-party system in the U.S., where Democrats are generally more favorable to immigrants than Republicans (Ollerenshaw & Jardina, 2023), may explain the public's polarization on the issue outlined by Jin (2024). However, Hainmueller and Hopkins (2015) found similarities in views about immigrants in the U.S., with both groups being favorable to highly educated immigrants with high-status employment and less favorable to those who arrived without authorization, lacked work plans, do not speak English, etc. Similarly, researchers have documented that higher-earning, highly educated individuals are more politically engaged than those who are not (Laurison, 2016).

Party preference influences attitudes toward immigration as individuals align their views with those of the party they identify with (Harteveld et al., 2017). Individuals' education levels and elite cues also shape public attitudes toward immigration (Hellwig & Kweon, 2016). Specifically, Hellwig and Kweon (2016) found that individuals with higher levels of education had more favorable attitudes toward immigration, and political elites significantly influenced immigration attitudes—particularly for highly educated individuals. Likewise, Hainmueller and Hopkins (2014) found that attitudes about immigrants are primarily related to concerns about the cultural impact rather than the economic impact. "Political participation refers to voluntary activities undertaken by the mass public to influence public policy, either directly or by affecting the selection of people who make policies" through activities such as voting, supporting or donating to campaigns, protesting, and more (Uhlaner, 2015, p. 504). However, for the purpose of this study, we focus on political engagement related to electing political leaders.

Our interest in legal immigration support, immigration saliency, and political engagement builds on elite cue and saliency theories (Brade et al., 2016; Brader et al., 2008; Lavorgna & Corda, 2024; Paul & Fitzgerald, 2021). Past work by Afzal (2021, 2022, 2024) shows how online exposure, legislative tone during crises, and economic hardship shape public opinion on immigration (Afzal, 2022, 2024; Bin Afzal, 2023). This study builds on that line by asking whether political engagement makes immigration feel more urgent or visible to U.S. voters. Doyle (2025) notes that policy options like skills-based visas exist, but public debate rarely reflects this nuance (Doyle, 2025). The prior literature and research show that what matters most is not whether immigration is present in the discourse but how it's framed and who pays attention. We operationalize the construct of public attention, engagement, and exposure to current policy discussions in their political spaces through a novel political engagement index (PAX). We explore further that partisanship remains the dominant influencing factor in shaping legal immigration saliency, and support or different socio-demographic factors mediate the influences.

## Methods:

We use the cross-sectional data from the 2024 ANES Pilot Study to explore how political engagement relates to public views on legal immigration. The dependent variable measures views on legal immigration policy, ranging from 1 ("A lot harder") to 5 ("A lot easier"). The independent variable is the Political Engagement Index (PAX), which sums up five binary indicators of civic activity: talking to others about politics, posting or sharing political content online, attending a rally or demonstration, wearing a political button, or displaying a sign, and volunteering for a campaign. The index ranges from 0 to 5. Table 1 outlines the construction of PAX for the study.

### Table 1. Political Engagement Index (PAX) Construction
*Measurement, Scale Reliability, and Factor Analysis (N = 1,721)*

| Component Activity | Mean | SD | Factor Loading (EFA) | CFA Loading | Unique Variance |
|---|---|---|---|---|---|
| Talked to others | .40 | .49 | 0.60 | 1.00† | .64 |
| Posted/shared online | .17 | .38 | 0.79 | 1.25 | .37 |
| Attended rally/demonstration | .13 | .33 | 0.79 | 1.16 | .38 |
| Wore button/sticker | .21 | .41 | 0.70 | 1.04 | .51 |
| Worked/volunteered for the campaign | .07 | .26 | 0.64 | 0.57 | .60 |

## Scale Reliability:

1. Cronbach's Alpha = **.731**
2. One-factor solution explains **50.2%** variance (EFA)
3. CFA fit: RMSEA = **0.061**, CFI = **0.983**, TLI = **0.966**, SRMR = **0.025**
4. **Note.** †First loading constrained to 1 in CFA for model identification.

We also include several controls. Respondent gender is coded as 1 for male and 0 for female. Age is logged to reduce skews and improve model fit. Education runs from 1 (less than high school) to 5 (graduate degree). Household income is measured on a 12-point scale, from under $10,000 to $150,000 and above. Two dummy variables capture self-identified Democrats and Republicans. Political independents serve as the baseline. Generalized social trust is measured on a 5-point scale, from "not at all" to "very much." Urban residence is a binary indicator of whether the respondent lives in a major city.

The principal analysis uses ordered logistic regression with survey weights applied to adjust for complex design and item non-response. Three models are estimated: a bivariate model, a complete model with controls, and a model with interaction terms between partisanship and urban residence. In addition to the full-sample sample, the analysis runs separate regressions for Democrats, Republicans, and Independents. We aim to assess whether the relationship between political engagement and immigration views varies by partisan identity. The Political Engagement Index ($\alpha$ = .731) shows good reliability ($\alpha$ = .731). A one-factor structure explains over 50% of the variance. Confirmatory factor analysis fits the data well, with strong loadings for each activity. All regressions use complete case analysis, with a final sample size of 1,357. Table 2 outlines the summary statistics for the current study and includes all the variables. We also apply complete case analysis and survey weights to ensure population-representative estimates and prevent bias from item nonresponse. All variables are listed in _Appendix A_, and _Appendix B_ confirms no multicollinearity concerns in the model, with all VIFs below 3.3, tolerances above 0.29, and a stable condition index. Model estimates remained stable and consistent across specifications, supporting the reliability of our ordered logistic results.

**Findings and Discussion:**

Table 2 shows the multilevel modeling that aims to capture how the cumulative Political Engagement Index (PAX) influences public views on legal immigration. In the bivariate model, each added activity in the Political Engagement Index increases the odds of supporting easier immigration by about 20% (OR = 1.201, p < .001). The effect stays positive and significant after adjusting gender, age, education, income, party ID, trust, and city size (Model 2: OR = 1.118, p < .05). Political engagement activities alone do not explain the outcome. However, it matters even after we control other variables.

**Table 2. Ordered Logistic Regression Models Predicting Views on Legal Immigration (DV)**
_Dependent Variable: Legal Immigration Views (1 = A lot harder, 5 = A lot easier)_

| Predictor | Model 1 (Bivariate) | Model 2 (With Controls) | Model 3 (With Interactions) |
|---|---|---|---|
| **Political Engagement** | 1.201 *** (0.053) | 1.118 * (0.045) | 1.112 * (0.047) |
| **Male** | — | 0.804 * (0.088) | 0.796 * (0.086) |
| **Log Age** | — | 0.305 *** (0.044) | 0.309 *** (0.044) |
| **Education** | — | 1.204 *** (0.050) | 1.204 *** (0.050) |
| **Income** | — | 0.998 (0.017) | 0.998 (0.017) |
| **Democrat** | — | 1.571 *** (0.204) | 1.577 ** (0.210) |
| **Republican** | — | 0.616 *** (0.086) | 0.561 *** (0.078) |
| **Social Trust** | — | 1.390 *** (0.087) | 1.388 *** (0.088) |
| **Big City** | — | 0.836 (0.123) | 0.696 (0.096) |
| **Democrat × Big City** | — | — | 1.107 (0.340) |
| **Republican × Big City** | — | — | 1.993 (0.808) |
| **Pseudo R²** | 0.006 | 0.064 | 0.065 |
| **AIC** | 4200.66 | 3971.91 | 3971.77 |
| **BIC** | 4226.73 | 4039.68 | 4049.97 |
| **Log Likelihood** | −2095.33 | −1972.95 | −1970.88 |
| **N** | 1,357 | 1,357 | 1,357 |

**Note:** Exponentiated odds ratios were reported; _p < .05 (*), p < .01 (**), p < .001 (***)._

We also observe in Model 2 that men are less likely than women to favor easier immigration (OR = 0.804, p < .05). Both Model 2 and Model 3 see the same trends regarding gender and preferences in legal immigration policy. Age also plays a significant role in shaping views about legal immigration policy (OR = 0.305, p < .001). Education and trust are associated with more supportive views toward legal immigration, and both are strong predictors across models two and three. Income and urban residence do not show substantial effects. Partisanship plays a clear role: Democratic respondents in our sample are more likely to support easier immigration policies (OR = 1.571, p < .001), while Republicans are less likely to support easier immigration (OR = 0.616, p < .001). Model 3 adds interaction terms for party and city size. The Republican–urban interaction (OR = 1.993, p = 0.07) points to a possible shift among city-dwelling Republicans, but the result does not reach standard significance levels.

Now, we explore the partisanship layers to observe whether self-reported partisanship influences public attitudes toward legal immigration stances. Table 3 breaks the analysis into party groups. The direction stays the same—more engagement predicts more support for legal immigration, but the estimates are smaller and not statistically substantial. Among Republicans, the effect is most significant (OR = 1.159), but the p-value is just above 0.10. Among Democrats and Independents, the effect is modest and not important.

**Table 3. Ordered Logistic Regression by Political Engagement (Subset by Party ID)**
*Dependent Variable: Legal Immigration Views (1 = A lot harder, 5 = A lot easier)*

| Predictor | (1) Democrats | (2) Republicans | (3) Independents |
|---|---|---|---|
| Political Engagement Index | 1.064 (0.070) | 1.159 (0.104) | 1.083 (0.107) |
| | | | |
| Male | 0.973 (0.183) | 0.858 (0.185) | 0.590 ** (0.119) |
| Log Age | 0.449 *** (0.108) | 0.265 *** (0.079) | 0.230 *** (0.060) |
| Education | 1.188 * (0.082) | 1.207 * (0.098) | 1.286 *** (0.096) |
| Household Income | 1.000 (0.029) | 1.009 (0.032) | 0.996 (0.033) |
| Social Trust | 1.372 ** (0.162) | 1.313 * (0.152) | 1.584 *** (0.181) |
| Lives in Big City | 0.802 (0.174) | 1.216 (0.353) | 0.733 (0.230) |
| | | | |
| Log Likelihood | −676.732 | −530.147 | −552.688 |
| AIC | 1375.464 | 1082.293 | 1127.375 |
| BIC | 1420.835 | 1125.836 | 1170.746 |
| Pseudo R² | 0.031 | 0.052 | 0.058 |
| N | 457 | 387 | 381 |

**Note:** Exponentiated odds ratios were reported; *p < .05 (\*), p < .01 (\*\*), p < .001 (\*\*\*)*.

We observe consistent patterns across the three self-identified partisan groups. Across all models, respondents with higher age values prefer less easy legal immigration policies. In contrast, higher formal education and generalized social trust are associated with more significant support for easing legal immigration rules. Among Independents, male respondents are notably less likely to favor more permissive immigration policies (OR = 0.590, p < .01), a pattern not observed in the Democratic or Republican subsets.

Our cross-sectional analysis of the ANES 2024 dataset shows that Political Engagement is positively associated with support for legal immigration across all partisan identities. However, when models are estimated separately by group, the strength and precision of this relationship weaken. This may reflect reduced statistical power in the smaller subsamples or differences in how engagement corresponds with policy preferences across partisan lines. The findings suggest that civic participation is modestly linked to more favorable attitudes toward legal immigration, though the effect is not uniform across all political affiliations.

## Predicted Probabilities by Political Engagement and Partisan Identity
### Views on Legal Immigration: A Lot Harder vs A Lot Easier

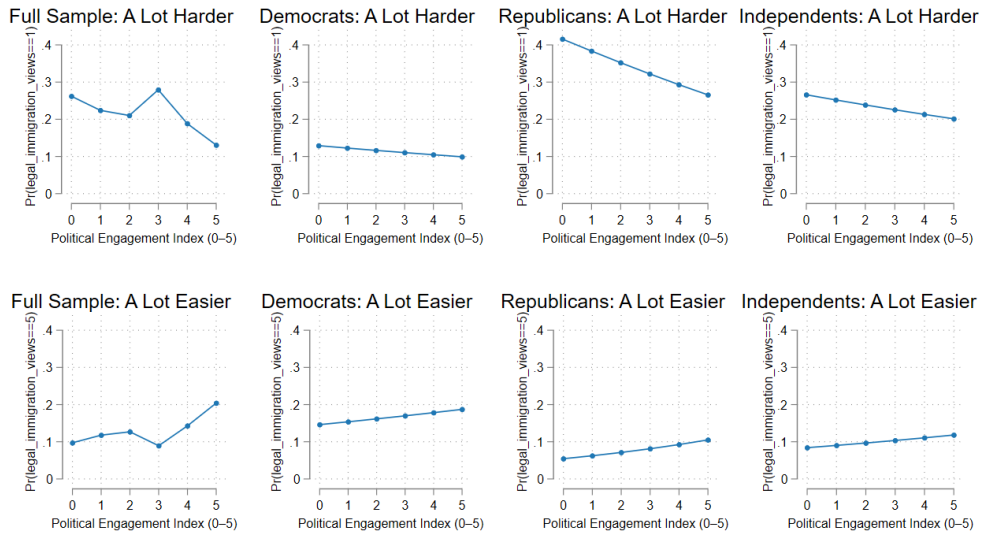

Figure 1: Predictive margins across party lines (Views on Legal Immigration and PAX)

We explore further how self-identified partisanship influences individuals' views on legal immigration concerning varying levels of political engagement. We visualize these trends in Figure 1, which predicted probabilities of choosing either "A lot harder" or "A lot easier" immigration policies across levels of political engagement. This figure offers an intuitive view of how engagement correlates with the most polarized policy attitudes in the full-sample model and partisan subgroups. Since the ordered logistic regression estimates can be difficult to interpret directly, plotting these probabilities allows us to assess better the practical effect of engagement, particularly at the policy extremes.

Table 4 builds on these visualizations by reporting the change in predicted probabilities between the lowest and highest levels of political engagement (0 to 5). The shifts in preferences about legal immigration policies help clarify the findings' substantive significance. We observe that, among Republicans, the likelihood of favoring much stricter immigration policies drops by 13 points (from 0.40 to 0.27), while support for making immigration "a lot easier" rises by 5 points. In contrast, Democrats show only modest change, suggesting that political engagement may have less additional influence among those already supportive of easier immigration policies. These differences across groups help explain why the strength and precision of engagement's effect diminish when the party estimates models separately.

**Table 4. Change in Predicted Probabilities Policy Preferences by Political Engagement Level[3]**

| Group | Outcome 1: "A Lot Harder" | Δ Probability | Outcome 5: "A Lot Easier" | Δ Probability |
|---|---|---|---|---|
| **Full Sample** | 0.26 → 0.13 | −0.13 | 0.10 → 0.21 | +0.11 |
| **Democrats** | 0.13 → 0.10 | −0.03 | 0.14 → 0.19 | +0.05 |
| **Republicans** | 0.40 → 0.27 | −0.13 | 0.07 → 0.12 | +0.05 |
| **Independents** | 0.27 → 0.20 | −0.07 | 0.10 → 0.13 | +0.03 |

We use predicted probabilities to show how civic engagement shapes public opinion on legal immigration. Table 5 highlights the change from low to high engagement across partisan groups. In the full-sample model, the probability of choosing "A Lot Harder" drops by 13 percentage points (from 0.26 to 0.13), while support for "A Lot Easier" rises by 11 points (from 0.10 to 0.21). These shifts are relatively smaller but still present in the party-specific models. Republicans show the same 13-point decline in hardline views but only a modest 5-point increase in more open views. Democrats show the smallest movement, with only a 3-point drop in restriction and a 5-point gain in support for easier policies. Independents fall in between, shifting 7 points away from "A Lot Harder" and 3 points toward "A Lot Easier."

---

[3] Predicted probabilities are estimated using weighted ordered logistic regression models. Probabilities represent the likelihood of selecting either the most restrictive (Outcome 1) or most permissive (Outcome 5) position on legal immigration at the lowest (0) and highest (5) levels of political engagement.

This analysis shows that cumulative political participation may meaningfully correspond with legal immigration attitudes. It shows how participation maps onto fundamental changes in opinion. Engagement links to more open immigration views, but the strength of this link depends on party identity. The results suggest that engagement alone does not produce uniform shifts. Instead, partisanship conditions are based on how civic activity aligns with immigration attitudes.

Figure 2 displays the predicted probabilities based on education levels and party affiliation. It indicates how education levels in the full-sample model and party affiliation relate to views on legal immigration. The model suggested that as the education level increases, the public generally leaned toward more lenient legal immigration policies. However, this effect was greater for Democrats and Independents than Republicans. We observe the predictive margins for full models and then use the sub-group analysis to observe the increase in attained education and how that impacts public opinion about legal immigration. The margin plots (figure 1) show that when the level of attained education increases, the directionality is consistent, and the attitude toward easier legal immigration also increases. We hold the PAX as constant in these margin plots and observe the pronounced effect only on education and views on the legal immigration process. While prior studies find that education is linked to more favorable immigration attitudes (Umansky et al., 2025; Hannuksela et al., 2024), our model holds political engagement constant, isolating education's independent effect on support for legal immigration.

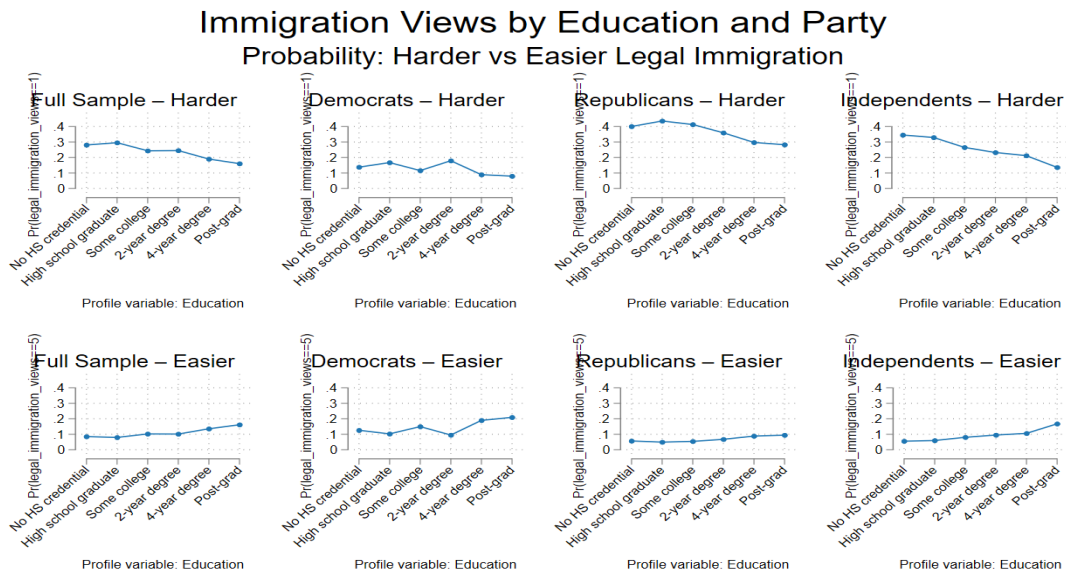

Figure 2: Predictive margins across education and party lines (Views on legal immigration)

Social trust, i.e., individuals' trust in others, also led to more favorability for legal immigration. As Figure 3 shows, the rate of change for Republicans was minimal compared to Democrats and Independents.

## Predicted Probabilities by Political Engagement and Partisan Identity
### Views on Legal Immigration: A Lot Harder vs A Lot Easier

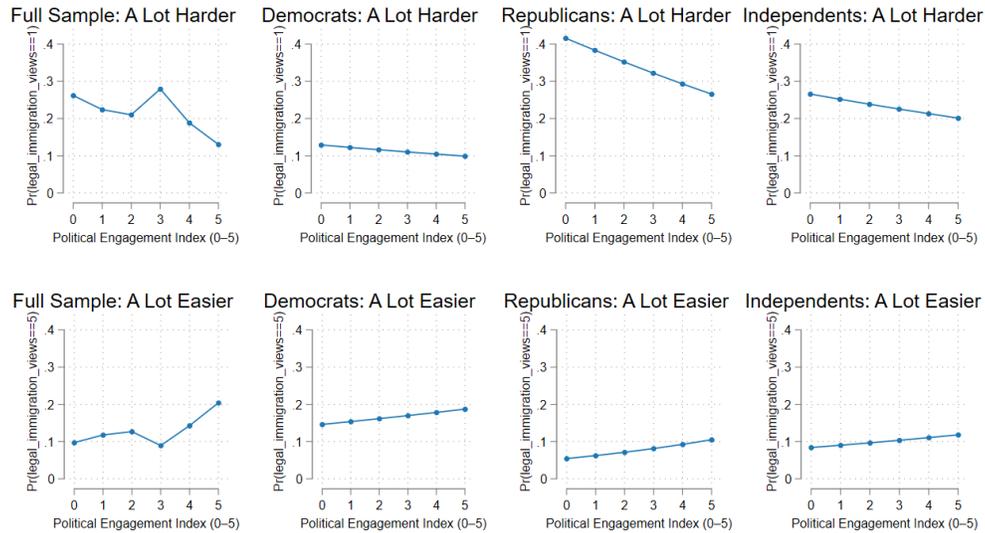

Figure 3: Predictive margins for Social Trust and Party (Views on legal immigration)

## Unique Contributions of the Study

This study makes several important contributions that extend our understanding of influences on attitudes toward legal immigration. First, our Political Engagement Index (PAX) was a reliable measure for gauging political participation and can be applied to other study contexts besides legal immigration. Secondly, our study calls for a more nuanced approach to understanding how political identity and demographics, such as age, gender, education, education, partisanship, and social trust, shape public opinions and attitudes regarding legal immigration. Although increased political engagement corresponded with increased support for legal immigration, there were noted differences based on party alignment. The effect of higher levels of political engagement was most pronounced among Republican respondents. In other words, political engagement alone does not shape individuals' immigration attitudes, as partisanship appears to moderate this relationship. Thirdly, we highlight the important role that education and social trust play in shaping immigration attitudes. Specifically, regardless of political affiliation, higher education levels were linked to more accepting views of legal immigration and higher levels of social trust. However, the effect was minimal with Republicans compared to other political affiliations.

## Limitations:

While this study offers timely insights into how cumulative political engagement activities shape attitudes toward legal immigration policies, our current study has limitations. First, the analysis draws on cross-sectional data from the 2024 ANES Pilot Study, which captures a single point in time. As a result, it does not establish causal direction, whether engagement drives opinion change or vice versa. Second, while the Political Engagement Index (PAX) demonstrates solid reliability, it captures self-reported behavior and may be subject to social desirability bias. Third, subgroup analyses by partisan identity reduce the sample size for each model, which may limit precision and obscure smaller but meaningful effects. Lastly, the study focuses on legal immigration as a single policy domain. It does not address whether these findings extend to attitudes on other forms of immigration or to broader policy domains. Despite these limitations, the findings underscore the need to take political behavior seriously when studying public opinion and suggest that civic participation plays a modest but consistent role in shaping immigration attitudes.

## Conclusion and Future Direction

This study examines how political engagement, education, and social trust shape public attitudes toward legal immigration in the United States. Prior research has linked education to more inclusive immigration views (Hainmueller & Hopkins, 2015; Umansky et al., 2025), social trust to tolerance and openness (Dinesen & Sønderskov, 2020; Rapp, 2024), and political engagement to ideological reinforcement or deliberative openness depending on context (Hannuksela et al., 2024; Kende et al., 2024). Our findings confirm some of these associations but depart from others: for instance, political engagement is significantly associated with more favorable immigration attitudes, particularly among Republican respondents. This contrasts with studies where engagement aligns only with liberal or centrist ideological profiles (Hannuksela et al., 2024).

Moreover, while prior research often shows education as a strong predictor of pro-immigration sentiment, we find a more complex relationship. In our sample, the effect of education is present but less pronounced than expected, echoing recent studies from Brazil, South Africa, and European countries that report no direct link between higher education and immigration tolerance (Kende et al., 2024; Umansky et al., 2025; Green & Staerklé 2023). This variation suggests that educational effects may be context-specific or moderated by other factors, such as political identity or social trust. On the other hand, our findings on social trust reinforce previous work showing its strong relationship with positive immigration views (Dinesen & Sønderskov, 2020) and extend this insight to the U.S. context using a full weighted sample.

A key contribution of this study is the utility and application of the Political Engagement Index (PAX), which combines behavioral indicators such as information-seeking, discussion, and participation. The PAX captures a broad form of democratic involvement, not limited to vote intention or partisanship, and offers a replicable framework for assessing engagement across topics. The utility of PAX could also be applied to evaluate voting behavior, issue saliency among the general population, and redevelop social policies more aligned with salient public needs and concerns. We also recommend that future collaborative research work may test its validity across other policy domains and political systems. Additionally, longitudinal research could better identify whether and how political engagement conditions change immigration attitudes. Finally, scholars should continue to examine how elite cues, media frames, and local contexts moderate the relationship between political engagement and immigration opinions, especially in polarized environments.

# **Appendices**

Appendix A. Variable Definitions and Distributions

### **1. Immigration Importance**
**What it captures:** Perceived importance of *illegal immigration* as a national issue.
**Usage:** It is not used in regression models but is included as a descriptive issue salience measure.

| Response | Frequency | Percent | Cumulative |
|---|---|---|---|
| Not Important | 62 | 4.01% | 4.01% |
| Slightly Important | 136 | 8.80% | 12.82% |
| Moderately Important | 267 | 17.28% | 30.10% |
| Very Important | 353 | 22.85% | 52.94% |
| Extremely Important | 727 | 47.06% | 100.00% |

### **2. Legal Immigration Views**
**What it captures:** Respondent's stance on how hard/easy legal immigration should be.
**Usage: Main DV** in all models (Model 1, 2, 3, and all subset regressions).

| Response | Frequency | Percent | Cumulative |
|---|---|---|---|
| A lot harder | 369 | 23.88% | 23.88% |
| Somewhat harder | 267 | 17.28% | 41.17% |
| About the same | 455 | 29.45% | 70.61% |
| Somewhat easier | 277 | 17.93% | 88.54% |
| A lot easier | 177 | 11.46% | 100.00% |

### **3. Political Engagement Index (0–5)**
**What it captures:** Composite index based on 5 political activities.
**Usage: Main IV** in all models and subset regressions.

| Score | Frequency | Percent | Cumulative |
|---|---|---|---|
| 0 | 763 | 49.42% | 49.42% |
| 1 | 379 | 24.55% | 73.96% |
| 2 | 177 | 11.46% | 85.43% |
| 3 | 98 | 6.35% | 91.77% |
| 4 | 88 | 5.70% | 97.47% |
| 5 | 39 | 2.53% | 100.00% |

### **4–8. Political Engagement Components (Binary Inputs)**
**Usage:** Not used independently in models but used to **build the index**.
### **4. Talked to others (mobil_talk)**

| Response | Frequency | Percent |
|---|---|---|
| No | 918 | 59.46% |
| Yes | 626 | 40.54% |

### **5. Participated online (mobil_online)**

| Response | Frequency | Percent |
|---|---|---|
| No | 1,265 | 81.88% |
| Yes | 280 | 18.12% |

**6. Attended rallies (mobil_rally)**

| Response | Frequency | Percent |
|---|---|---|
| No | 1,311 | 84.85% |
| Yes | 234 | 15.15% |

**7. Wore button/sticker (mobil_button)**

| Response | Frequency | Percent |
|---|---|---|
| No | 1,208 | 78.19% |
| Yes | 337 | 21.81% |

**8. Worked/volunteered (mobil_work)**

| Response | Frequency | Percent |
|---|---|---|
| No | 1,448 | 93.72% |
| Yes | 97 | 6.28% |

**9. Gender**

**What it captures:** Binary gender recode.

**Usage:** Control variable in all models.

| Gender | Frequency | Percent |
|---|---|---|
| Female | 811 | 52.49% |
| Male | 734 | 47.51% |

**10. Log of Age**

**What it captures:** Age transformed using natural log.

**Usage:** Control in all models.

| Statistic | Value |
|---|---|
| Obs | 1,545 |
| Mean | 3.84 |
| Std. Dev. | 0.39 |
| Min | 2.89 |
| Max | 4.54 |

**11. Education**

**What it captures:** Highest educational attainment.

**Usage:** Control in all models and subset margins.

| Education Level | Frequency | Percent |
|---|---|---|
| No HS credential | 51 | 3.30% |
| High school graduate | 464 | 30.03% |
| Some college | 310 | 20.06% |
| 2-year degree | 174 | 11.26% |
| 4-year degree | 355 | 22.98% |
| Post-graduate | 191 | 12.36% |

**12. Income**

**What it captures:** Household income brackets.

**Usage:** Control in all models.

| Income Range | Frequency | Percent |
|---|---|---|
| <$10K | 103 | 6.67% |
| $10K–$19K | 115 | 7.44% |
| $20K–$29K | 203 | 13.14% |
| $30K–$39K | 126 | 8.16% |
| $40K–$49K | 127 | 8.22% |
| $50K–$59K | 127 | 8.22% |
| $60K–$69K | 106 | 6.86% |
| $70K–$79K | 135 | 8.74% |
| $80K–$99K | 138 | 8.93% |
| $100K–$119K | 117 | 7.57% |
| $120K–$149K | 101 | 6.54% |
| $150K+ | 147 | 9.51% |

**13. Generalized Social Trust**

**What it captures:** Trust in people in general.

**Usage:** Included as a zinger control variable in all models.

| Trust Level | Frequency | Percent |
|---|---|---|
| Never trust others | 125 | 8.10% |
| Some of the time | 485 | 31.41% |
| About half the time | 430 | 27.85% |
| Most of the time | 492 | 31.87% |
| Always trust others | 12 | 0.78% |

**14. Big City Residence**

**What it captures:** Urban vs. non-urban location.

**Usage:** Contextual control and interaction term.

| Response | Frequency | Percent |
|---|---|---|
| No | 1,248 | 80.78% |
| Yes | 297 | 19.22% |

**15–17. Party ID (Moderator Dummies)**

**What it captures:** 3 mutually exclusive dummies.

**Usage:** Controls and subset identifiers in all models.

| Party ID | Yes (%) | No (%) |
|---|---|---|
| Democrat | 32.94% | 67.06% |
| Republican | 28.93% | 71.07% |
| Independent | 27.06% | 72.94% |

**18. Interaction Terms**

**What they capture:** Conditional effects of party ID × political engagement or residence.

**Usage:** Only in Model 3 and relevant subsets.

**Democrat × Big City**

| Value | Freq | % |
|---|---|---|
| 0 | 1,385 | 89.64% |
| 1 | 160 | 10.36% |

**Republican × Big City**

| Value | Freq | % |
|---|---|---|
| 0 | 1,493 | 96.63% |
| 1 | 52 | 3.37% |

**Appendix B. Multicollinearity Diagnostics (All Models)**

| Test | Result | Note |
|---|---|---|
| VIF (Mean / Max) | 1.68 / 3.29 | All < 5 |
| Tolerance (Min) | 0.2991 | All > 0.1 |
| Pairwise Correlation Max | 0.70 (big_city × dem_bigcity) | No bivariate multicollinearity |
| Condition Index (Max) | 41.06 | Acceptable with low VIFs |
| Correlation Matrix Determinant | 0.0880 | No singularity |

**Appendix C. Coefficient Plot (All Three Multilevel models using full dataset)**

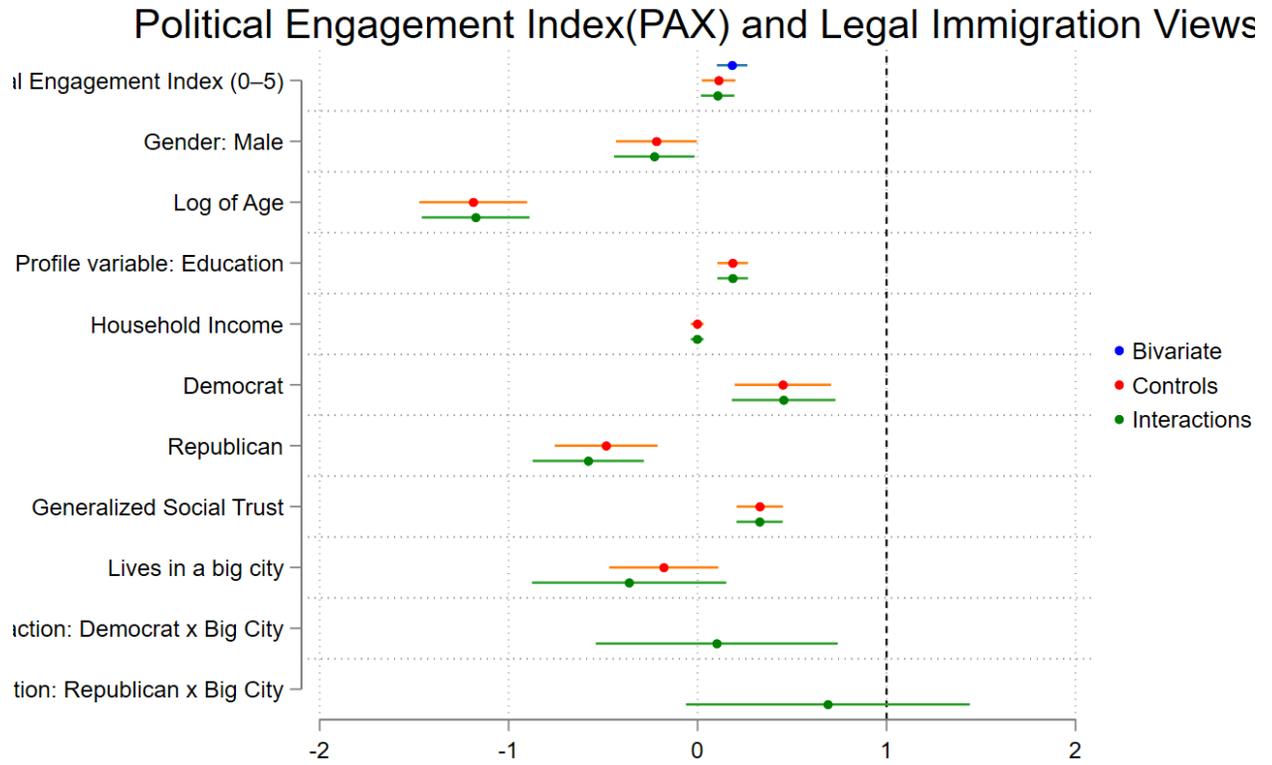

Figure: The coefficient plot full dataset (ANES, 2024)

**Appendix D. Coefficient Plot (All Three Multilevel models focusing on Party ID)**

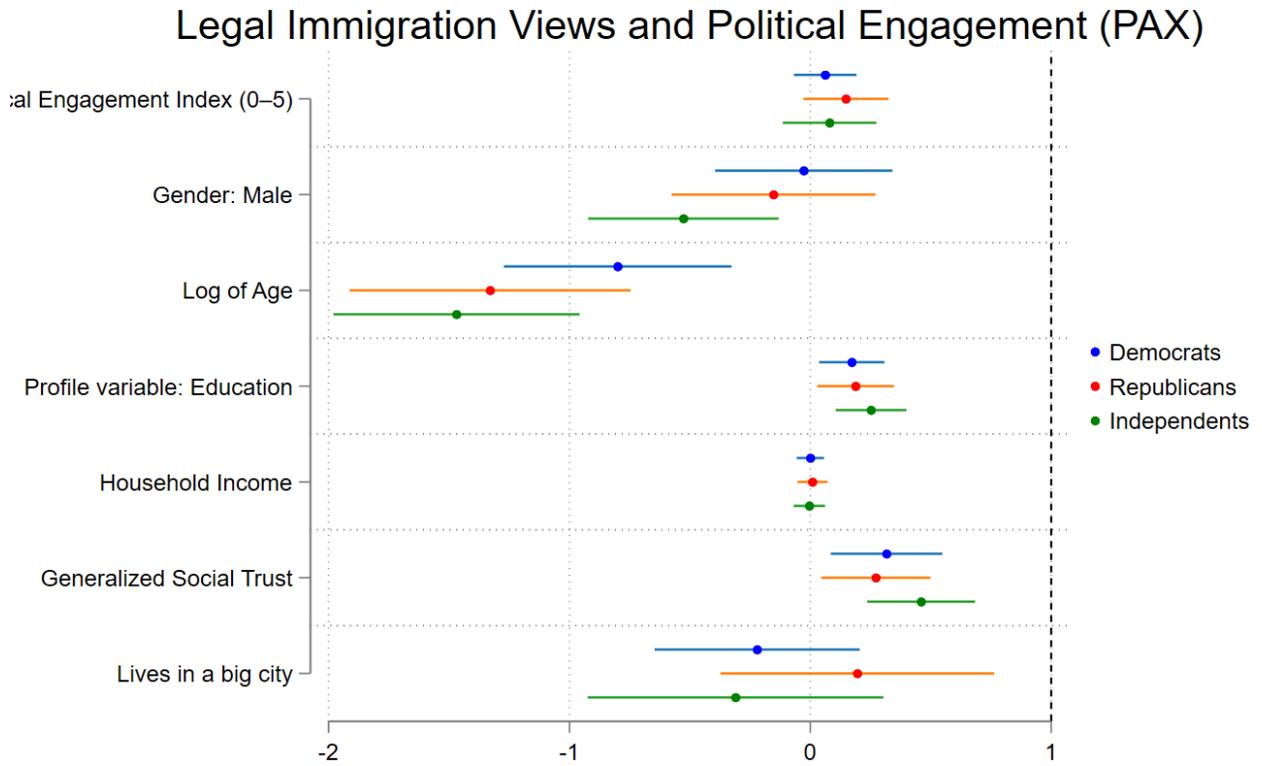

Figure: The coefficient plot based on self-reported Party ID (ANES, 2024)

**Appendix E. Change in Predicted Probabilities by Education Level**

**Table:  Change in Predicted Probabilities by Education Level**

| Group | Outcome 1: "A Lot Harder" | Δ Probability | Outcome 5: "A Lot Easier" | Δ Probability |
|-------|---------------------------|---------------|---------------------------|---------------|
| **Full Data** | 0.26 → 0.13 | **−0.13** | 0.10 → 0.21 | **+0.11** |
| **Democrats** | 0.14 → 0.10 | **−0.04** | 0.12 → 0.18 | **+0.06** |
| **Republicans** | 0.41 → 0.27 | **−0.14** | 0.06 → 0.09 | **+0.03** |
| **Independents** | 0.37 → 0.19 | **−0.18** | 0.08 → 0.18 | **+0.10** |

**Note: Profile Variable:** *Education;* **Outcome:** *Views on Legal Immigration;* (*Predicted Probability Difference: From "No HS Credential" to "Post-grad"*)

**Appendix F. Change in Predicted Probabilities by Generalized Social Trust**

**Table: Change in Predicted Probabilities by Generalized Social Trust**

| Group | Outcome 1: "A Lot Harder" | Δ Probability | Outcome 5: "A Lot Easier" | Δ Probability |
|-------|---------------------------|---------------|---------------------------|---------------|
| **Full Data** | 0.45 → 0.14 | **−0.31** | 0.10 → 0.31 | **+0.21** |
| **Democrats** | 0.53 → 0.11 | **−0.42** | 0.07 → 0.37 | **+0.30** |
| **Republicans** | 0.47 → 0.29 | **−0.18** | 0.07 → 0.09 | **+0.02** |
| **Independents** | 0.44 → 0.15 | **−0.29** | 0.07 → 0.27 | **+0.20** |

**Note: Profile Variable:** *Generalized Social Trust;* **Outcome:** *Views on Legal Immigration;* (*Predicted Probability Difference: From "Never Trust" to "Always Trust"*)